Università dell'Aquila
Dipartimento di Ingegneria Elettrica

**Multimethods and separate static typechecking in a language with C++-like object model**

Emanuele Panizzi, Bernardo Pastorelli

Research report R.99-33

# Multimethods and separate static typechecking in a language with C++-like object model

Emanuele Panizzi, Bernardo Pastorelli
{panizzi, pastorelli}@ing.univaq.it

Università degli Studi dell'Aquila, Dipartimento di Ingegneria Elettrica
67040 Monteluco di Roio, L'Aquila, Italy

**Abstract**

The goal of this paper is the description and analysis of multimethod implementation in a new object-oriented, class-based programming language called OOLANG. The implementation of the multimethod typecheck and selection, deeply analyzed in the paper, is performed in two phases in order to allow static typechecking and separate compilation of modules. The first phase is performed at compile time, while the second is executed at link time and does not require the modules' source code. OOLANG has syntax similar to C++; the main differences are the absence of pointers and the realization of polymorphism through subsumption. It adopts the C++ object model and supports multiple inheritance as well as virtual base classes. For this reason, it has been necessary to define techniques for realigning argument and return value addresses when performing multimethod invocations.

## 1. Introduction

When writing object-oriented mathematical software it would be useful to define a polymorphic binary operator in a base class and in a class derived from it in such a way that:

- each definition accept two arguments (e.g. the invocation object and a parameter) of the same type of the class to which it belongs;
- the proper definition be called according to the runtime type of both the operands.

This is generally not possible in traditional object-oriented languages because they use single dispatching techniques, so only the type of one operand (the invocation object) is used to select the definition to call.

One of the solutions to this problem (also known as the problem of binary methods [7]) is the use of *multimethods* [7][19]. They are polimorphic functions selected considering not only the type of the invocation object but also the type of all the other arguments.

This solution has been implemented in OOLANG, a new class-based object-oriented language targeted at mathematical software[1].

For performance reasons, one of the goals of OOLANG is the ability to perform typechecking at compile-time (i.e. to support static typechecking). Moreover, in order to support precompiled libraries, OOLANG allows separate compilation. But it's surely challenging the separate static typechecking of multimethods: in fact, as multimethods can be declared and defined outside class declarations (and eventually in different modules) two source files that typecheck successfully when compiled (separately) may interfere and generate type errors when linked together. In other languages that support multimethods this problem is solved either by requiring that the typechecking of the entire program be done in a whole or by imposing restrictions on the symmetry of multimethods. OOLANG, instead, allows the separate static typechecking of arbitrary multimethods, requiring at link-time only a simple check that is efficiently performed during the construction of the compressed dispatch tables [18].

Moreover, OOLANG implements multimethods over the C++ object model. This opens two problems due to the representation of objects that are instances of classes with multiple parents. The first is the loss (to a certain extent) of transitivity in the subtype relation. The second is the necessity to realign objects when

---

[1] In fact, OOLANG was developed for the APEmille SPMD supercomputer [2][4]. In this paper only the object-oriented features are analyzed, avoiding the description of the parallel constructs, not relevant to the argument of the paper.

```
class Point {
    int x,y;
    virtual void dump(){
        printf("Point\n");
    }
};

class ColorPoint: public Point {
    int color;
    void dump(){
        printf("ColorPoint\n");
    }
}

void print(Point p) {
    p.dump();
}

int main() {
    Point a;
    ColorPoint b;

    print(a);
    print(b);
};
```

(a)

---

```
Point
ColorPoint
```

(b)

**Figure 1. Example of OOLANG program (a) and its output (b)**

passed as arguments to a multimethod or when they are returned by multimethods to static functions.

The OOLANG language does not support pointers. It however supports subsumption, i.e. the possibility to use an instance of a derived class where an instance of a parent class is expected. Parameters (of a function or multimethod) can be passed by value (the whole object passed as argument is copied on the stack) or by reference (&). It is also possible to declare a parameter as constant and, to allow this, it has been necessary to solve a problem of interference of constant parameters with multimethod selection. Finally, because of the impossibility to know which specialization of a multimethod will be invoked at run-time and in order to effectively take advantage of the different parameter passing possibilities, it has been necessary to develop a parameter passing scheme that requires the cooperation of the caller and the callee.

The OOLANG compiler produces APEmille code as well as portable C code.

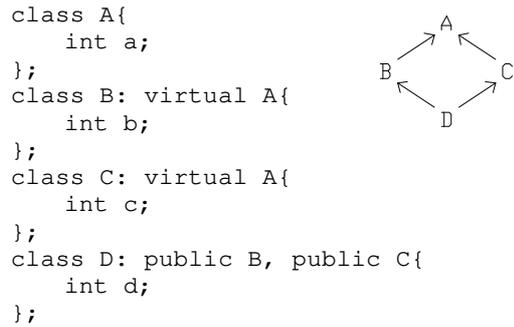

```
class A{
    int a;
};
class B: virtual A{
    int b;
};
class C: virtual A{
    int c;
};
class D: public B, public C{
    int d;
};
```

**Figure 2. Virtual inheritance**

The paper is organized as follows: section 2 introduces the main features of the OOLANG language. Section 3 discusses the OOLANG object model, pointing out the loss of transitivity in the subtype relation and the necessity of realigning objects. Section 4 presents the details of OOLANG typechecking of multimethods. Section 5 describes the mechanism of parameter passing and address realignment. Section 6 analyzes the related work while section 7 drains the conclusions and presents some future developments.

## 2. Introduction to the OOLANG language

OOLANG is a class-based object-oriented language with syntax similar to the C++ one [25]. The most significant OOLANG characteristics are shown in this section.

A first OOLANG feature is the support for virtual functions, i.e. functions that are selected at runtime considering the type of the invocation object (similar to those of C++). The main difference is that, unlike C++, OOLANG is pointer-less and allows polymorphism through subsumption, i.e. through the possibility of using an instance of a derived class where an instance of a parent class is expected[2]. Figure 1.a presents a fragment of OOLANG code[3]: the class *ColorPoint* inherits the *x* and *y* fields from the class *Point* (which is a public parent of *ColorPoint*). The static function *print()*, that accepts a *Point* argument, is invoked once using an argument of type *Point* and then using an argument of type *ColorPoint*. The *print()* function in turn invokes the virtual function *dump()*. Although the static type of the parameter *p* is *Point* and there

---

[2] C++ on the other hand allows polymorphism only through pointers, e.g. when the invocation object of a virtual function is a pointer.
[3] The code showed in the following figures is written in OOLANG language unless differently specified.

```
class Point {
    int x,y;
    virtual bool equal(Point p){
        return (x==p.x)&&(y==p.y);
    }
};

class ColorPoint: public Point {
    int color;
    virtual bool equal(Point p){
        return (x==p.x)&&(y==p.y);
    }
};
```

**Figure 3. Limitations of virtual functions**

are no pointers (as the language is pointer-less), thanks to subsumption, the invoked `dump()` function is the one defined in the class of the actual parameter (`A` or `B` respectively) as shown by the program output (Figure 1.b).

Another OOLANG characteristic is the support of multiple inheritance and virtual parents: if a class `D` inherits from two classes `B` and `C` that have a common parent `A` declared virtual (Figure 2), its objects contain only one copy of the fields of the `A` parent. This topic will be further analyzed in section 3.

The main OOLANG feature that will be extensively analyzed in this paper is the implementation of multimethods. Multimethods allow avoiding a fundamental limitation of virtual functions.

Figure 3 presents a new version of the `Point` and `ColorPoint` classes containing a virtual function `equal()` that tests if two points are equal. Both the versions of the `equal()` function have a parameter of type `Point`. This example (adapted from [7]) shows the cited limitation of virtual functions. It is not possible to define, in the derived class, a version of the `equal()` function that accepts an instance of the derived class as parameter: the parameters of a virtual function can't be specialized when the function is overridden in a new class.

If this constraint is relaxed, run-time type errors can arise. In fact, suppose that the `equal()` function defined in `ColorPoint` class accept an instance of this class as parameter (as in Figure 4, whose code is not good OOLANG). The invocation of function `func()` is correct because the `ColorPoint` object `a` is subsumed to an instance of `Point`. The invocation of the `equal()` method in function `func()` typechecks at compile-time because it is checked against the `equal()` method declared in class `Point`, having `p1` static type `Point`. At run-

```
class Point {
    int x,y;
    virtual bool equal(Point p){
        return (x==p.x)&&(y==p.y);
    }
};

class ColorPoint: public Point {
    int color;
    virtual bool equal(ColorPoint p){
        return (x==p.x)&&(y==p.y)&&
            (color==p.color);
    }
};

bool func(Point p1, Point p2){
    return p1.equal(p2);
}

int main(){
    ColorPoint a;
    Point b;

    return func(a,b);
}
```

**Figure 4. Problems with the specialization of virtual functions parameters. The code presented is not good OOLANG.**

time, the dynamic type of `p1` is `ColorPoint` because the first argument of `func()` is the `a` object; so the `equal()` method declared in class `ColorPoint` is invoked. This method expects a second argument of type `ColorPoint` while `p2` has static and dynamic type `Point` (it is a copy of the `b` object): this results in a run-time error because an instance of `Point` hasn't the `color` field that is accessed in the code of the method (this error, in the worst case, *does not* raise an exception).

To solve this problem OOLANG uses multimethods. These functions are selected considering the types of all the arguments of a method invocation and not only the type of the first one (the invocation object). OOLANG treats all the operators (for example `operator==` or `operator+`) like multimethods. In fact mathematical operators are the classical examples of binary methods and they are the primary reason of the inclusion of multimethods in OOLANG. The programmer can define other multimethods in the same way he defines ordinary methods; the only difference is that a multimethod name begins with `@`[4].

---

[4] In the present version of OOLANG if a function is declared to be a multimethod prefixing its name with @, all of its parameters of user-defined type will be used during the selection. In fact the programmer can't decide which parameters of a multimethod have to be used.

```
class Point {
    int x,y;
    bool @equal(Point p){
        return (x==p.x)&&(y==p.y);
    }
};

class ColorPoint: public Point {
    int color;
    bool @equal(ColorPoint p){
        return (x==p.x)&&(y==p.y)&&
            (color==p.color);
    }
};

bool func(Point p1, Point p2){
    return p1.@equal(p2);
}

int main(){
    ColorPoint a;
    Point b;

    return func(a,b);
}
```

**Figure 5. Use of multimethods in OOLANG to solve the binary method problem.**

```
class Point {
    int x,y;
    bool @equal(Point &p){
        return (x==px)&&(y==p.y);
    }
};

class ColorPoint: public Point {
    int color;
};

bool @equal(ColorPoint &p1,
        ColorPoint &p2) {
    return (p1.x==p2.x)&&(p1.y==p2.y)&&
        (p1.color==p2.color);
}

int main() {
    Point p;
    ColorPoint cp;

    @equal(p,cp);
    return cp.@equal(p);
}
```

**Figure 6. Examples of multimethod**

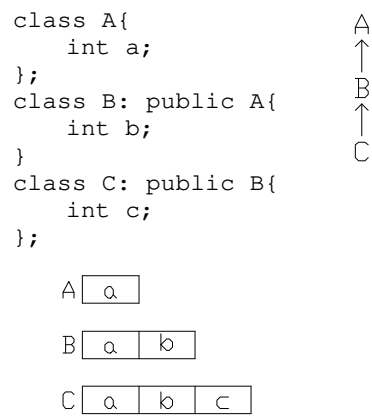

```
class A{
    int a;
};
class B: public A{
    int b;
}
class C: public B{
    int c;
};
```

**Figure 7. Sample class hierarchy and object layout**

In the same manner as virtual functions can be defined once for every class accepted as invocation object, multimethods can have a different definition for each combination of types of their arguments. In the following every single definition is referred to as *specialization* of the multimethod, while the word *multimethod* is used to refer to the collection of specializations without emphasis on a particular one[5].

An example of multimethod is reported in Figure 5. It is the `@equal()` multimethod and has two specializations: the first accepts two `Point` arguments while the second accepts two `ColorPoint` arguments (including the invocation object). As the two arguments `p1` and `p2` in `func()` have run-time types `ColorPoint` and `Point` respectively, the only applicable specialization of the multimethod is the first one (i.e. `Point::@equal(Point &)`). Thus that specialization will be invoked although the type of the invocation object is `ColorPoint` (it will be converted to `Point`).

A specialization of a multimethod can be declared and defined inside or outside a class (Figure 6). For example, the `@equal()` specialization defined inside the `Point` class could be equivalently defined outside using `@equal(Point &this, Point &p)`. In fact a multimethod defined inside a class is treated internally as a multimethod that has the same parameters plus, as first parameter, a reference (`&`) to the class type (`Point` in the example).

The invocation of a multimethod can be done using the syntax for a function invocation (as in the first

---

[5] The built-in types (integer, float, …) are not used during run-time selection but they are used to select the multimethod at compile-time. In fact, for example, *@m(int,Point)* and *@m(float,ColorPoint)* are not two specialization of the same multimethod but are two different multimethods with the same name and number of parameters but a different built-in type as first parameter. At compile-time, during an invocation *@m(a,b)* the first or the second multimethod is chosen based on the fact that *a* has type *int* or *float*.

```
class A {…};
class B {…};
class C {…};
class D: public A, public B {…};
class E: public C, public D {…};
```

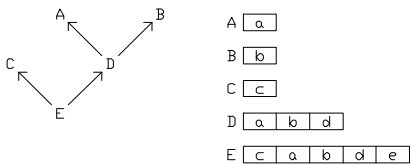

**Figure 8. Class hierarchy and object layout in presence of multiple inheritance**

```
class A{
    int a;
};
class B: public A{
    int b;
};
class C: public A{
    int c;
};
class D: public B, public C{
    int d;
};
```

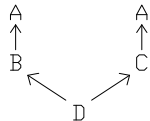

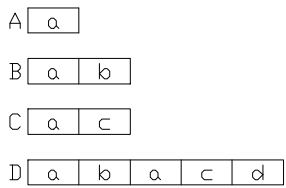

**Figure 9. Class hierarchy and object layout in presence of multiple inheritance**

case in the function `main()` or the method invocation syntax (as in the latter statement of `main()`). This second invocation syntax, together with the possibility of defining specializations of multimethods outside classes, allows to extend a class without modifying its source code and to invoke the new specializations as if they were methods of the class. For example, in Figure 6, `@equal()` is added to the class `ColorPoint` without modifying its source code and it is then invoked as if it were a member of the class.

```
class A{
    int a;
};
class B: virtual A{
    int b;
};
class C: virtual A{
    int c;
};
class D: public B,
public C{
    int d;
};
```

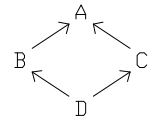

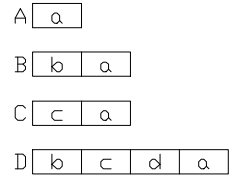

**Figure 10. Class hierarchy and object layout in presence of multiple inheritance and virtual parent**

Finally, a multimethod specialization is said to be *more specific* than another if the type of each parameter of the former is either the same type, or a derived type, of the corresponding parameter of the latter.

## 3. The OOLANG object model

The object layout in the OOLANG language has been chosen taking into account the time and space efficiency required by the applications for which the language has been developed, as well as the multiple inheritance support. Among the different possibilities found in literature [13][17][24], the C++ one has been chosen because it allows direct access to object fields (i.e. with a single memory access), prevents the creation of unused space in objects and supports multiple inheritance.

According to this object model, fields inherited from parents are located at the beginning of the object layout while fields relative to the object's class are at the end. Thus an object of a derived class is built "recursively" appending the fields of its class to the fields inherited from its base classes which have been set up in the same way. So each group of fields related to a base class is organized with the same layout as in its original class, setting up, in this way, a subobject (Figure 7).

In this model, when there is no multiple inheritance, each subobject starts at the beginning of the host object. In case of multiple inheritance, on the other hand, there are some subobjects that will not start at the beginning of the host object (e.g. subobject D in object E in Figure 8). We show in the following that this leads to the necessity of realigning the pointers generated by the compiler behind the scene in case of subsumption or multimethod invocation.

Due to the object model adopted, some ambiguities may arise in case of subsumption. In fact, when a class *D* (like in Figure 9) inherits from two classes *B* and *C* and both these classes inherit from the same class *A* (without virtual derivation), an instance of *D* contains two subobjects related to the *A* parents. In fact, a subobject *A* is contained both in the subobject *B* and in the subobject *C*. This leads to an ambiguity when trying to subsume an instance of *D* into an instance of *A*. It is important to stress that in the example above the transitivity of the subtype relation has been violated. In fact:

$$\begin{array}{l} D \leq B \text{ and } B \leq A \\ D \leq C \text{ and } C \leq A \end{array} \text{ but } \neg(D \leq A)$$

(where ≤ is used to indicate subtype relation). *D* is an ambiguous subtype of *A* and an instance of *D* cannot be converted to an instance of *A*.

To avoid the ambiguity in subtype relation it is possible to declare a parent class as virtual (Figure 10) as explained in section 2. When a class *D* inherits from other classes *B* and *C* that have the same virtual parent *A*, only one subobject relative to *A* is included in every instance of *D*. So an instance of *D* can be converted to an instance of *A* without ambiguity. Moreover, the subobject relative to a virtual parent is located at the end of the whole object (Figure 10).

OOLANG uses tables to dispatch virtual functions, to maintain pointers to base virtual classes and to maintain information needed during the realignment of return types. The first two tables are similar to those used by C++, so the reader can refer to [20] for further information. The third table (RTTABLE) is described below.

The RTTABLE for an object *o* contains:

- the id of the type associated to *o* (needed for multimethod selection, see section 4)
- the size of *o* (used for the copy of the object on the secondary stack, see section 5)
- the offset of the subobject from the beginning of the host object (the offset is 0 if *o* is a complete object)
- the number of parents of the class associated to *o*
- for each parent, the couple `(p_id,p_off)` where `p_id` is the parent id and `p_off` is the offset of the parent subobject from the beginning of the host object (*o*)

The last two pieces of information are necessary to realign an object to one of its parents (see section 5.3) while the first three are necessary for multimethod selection and parameter passing.

## 4. Multimethod typechecking and selection

Due to the fact that the OOLANG language allows to declare multimethod specializations outside class declarations (and eventually in different modules), it has been necessary to conceive a typechecking and selection mechanism divided into two phases.

The first phase takes place during compilation and takes care of the typecheck of multimethod invocations. For each invocation, at least an applicable specialization must be granted and a static return type is recognized. The impossibility of finding an applicable specialization is reported as error and stops the compilation, as described in the next section. On the other hand, any inconsistency or conflict among specialization declarations detected during this phase is only reported as a warning, because it could be solved by declarations made in other modules.

The second phase takes place at link time, is integrated in the generation of the compressed dispatch table needed for the runtime multimethod selection and doesn't require access to the source code of functions or multimethods. This is called *pre-link* phase and is needed to ensure that no conflicts or inconsistencies among multimethod specializations arise from the integration of different modules.

It is interesting to note that all the source code is typechecked during the first phase. The only problem that is checked at link time is the consistency of multimethods hierarchies, i.e. the absence of anomalies in the selection of the proper specialization and the satisfaction of a constraint on the return types of specializations defined in different modules.

These checks warrant that no error can occur at runtime due to multimethod management.

### 4.1 The first phase of multimethod type checking

When compiling code containing multimethods, it is necessary to check each multimethod invocation in

```
class A { … };                          class A { … };
class B { … };                          class B { … };
class C: public A, public B { … };      class C: public A, public B { … };

A @m(A,A);                              A @m(A,A);
B @m(B,B);                              B @m(B,B);
int f1(A);                              int f1(A);

int f2(C o1, C o2){                     int f2(A o1, A o2){
    return f1(@m(o1,o2));                   return f1(@m(o1,o2));
}                                       }
                    (a)                                     (b)
```

**Figure 11. Influence of anomalies on the determination of the return type of a multimethod**

order to establish if there exist at least an applicable specialization and to calculate the static return type of the invocation.

Three kinds of anomalies can occur during compilation:

a) no multimethod specialization is applicable to a particular multimethod invocation. Obviously it is possible that such a specialization be available in another source file, but if it is not available in the module under compilation then it is not possible for the compiler to type the multimethod invocation. It is thus necessary to stop the compilation process. Of course, a mere declaration of the proper multimethod specialization (without its definition) would be sufficient in order to avoid such compilation error.

b) no most-specific specialization exists. These situations are not necessarily errors. In fact an error is reported only if there exists an invocation that requires to be checked using the conflicting specializations. In the case of Figure 11.a, for example, it is not possible to establish the return type of the multimethod invocation because both the specializations are applicable and it is not possible to type the invocation. In this case it is necessary to report an error and stop the compilation because it is not possible to establish if the call to function `f1()` is correct. In the case of Figure 11.b, the `f2()` function can be typechecked without problems and thus the ambiguity problem is only reported as a warning and the compilation is not aborted. In fact the linking of different modules can create ambiguities but can also solve them: supposing that a module contains a specialization `@m(C,C)`, the ambiguity in Figure 11.b will be removed at link time. So if ambiguity doesn't interfere with the compilation process no error is reported and every decision is deferred at link time.

c) the last kind of anomaly can arise because of the ambiguity in subtype relation. For example considering the class hierarchy of Figure 9 and a multimethod specialization `@m(A,A)`, an invocation `@m(D,D)` will rise the anomaly because `D` is an ambiguous subtype of `A`. But in this situation only a warning is reported at compile-time because the return type of the invocation `@m(D,D)` is calculable. This kind of anomaly cannot interfere with the compilation process and could eventually be resolved at link-time due to other declarations present in different modules.

Therefore, during the first phase of compilation, an error is reported only if the anomaly interferes with the compilation process, preventing the calculation of the static type of a multimethod invocation (i.e. the return type of the multimethod that is statically foreseeable). In all the other cases, the anomaly generates a warning because it is possible that the union of more modules removes it.

To warrant soundness of static typing, OOLANG imposes on the return type of multimethods the same constraint presented in [3]: if a specialization of a multimethod is more specific than another, the return type of the former must be a subtype of the return type of the latter (in order to be accepted in expressions where the return type of the less specific specialization was statically expected). Moreover, because of ambiguity in the subtype relation, it is necessary to verify that the return type is not an ambiguous subtype of the return type of the less specific specialization. This check, that is made difficult by the loss of transitivity in the subtype relation, will be better analyzed in section 4.3.

| module classes.h:<br><br>```<br>class A {...};<br>class B: public A {...};<br>```<br><br>first source file:<br><br>```<br>#include "classes.h"<br>int @m(A a,B b){...}<br>int f() {<br>    B b1,b2;<br>    return @m(b1,b2);<br>}<br>```<br><br>second source file:<br><br>```<br>#include "classes.h"<br>int @m(B b,A a){...}<br>int main() {<br>    B b1,b2;<br>    return @m(b1,b2);<br>}<br>```<br><br>(a) | module classes.h:<br><br>```<br>class A {...};<br>class B: public A {...};<br>```<br><br>first source file:<br><br>```<br>#include "classes.h"<br>int @m(A a1,A a2){...}<br>int @m(B b1,B b2){...}<br>```<br><br>second source file:<br><br>```<br>#include "classes.h"<br>class C: public A, public B<br>{...}<br><br>int main() {<br>    C c1,c2;<br>    ...<br>    return @m(c1,c2);<br>}<br>```<br><br>(b) | module classes.h:<br><br>```<br>class A {...};<br>class B: public A {...};<br>class C: public A {...};<br>```<br><br>first source file:<br><br>```<br>#include "classes.h"<br>int @m(A a1,A a2){...}<br>```<br><br>second source file:<br><br>```<br>#include "classes.h"<br>class D: public B, public C<br>{...}<br><br>int main() {<br>    D d1,d2;<br>    ...<br>    return @m(d1,d2);<br>}<br>```<br><br>(c) |
|---|---|---|

**Figure 12. The three static typechecking difficulties arising in presence of separate compilation**

## 4.2. Problems of separate compilation

Static typechecking of arbitrary multimethods is challenging due to possible interference arising in the presence of multiple modules (i.e. source files). Source files that typecheck when compiled separately, can create ambiguity when linked together. In particular three ambiguous situations (three anomalies) can arise:

1) in Figure 12.a each of the two modules typechecks separately, but when they are linked, the invocation of `@m` becomes ambiguous because neither of the specialization is more specific than the other;

2) in Figure 12.b the ambiguity arises because of multiple inheritance. `C` inherits from `A` and from `B` so both the specializations declared in the first module are applicable to an invocation with two instances of `C` as arguments, but neither is more specific;

3) the last kind of ambiguity arises because of the loss of transitivity of the subtype relation. In Figure 12.c the definition of class `D` in the second module creates an ambiguity because an invocation of `@m` with run-time types `(D,D)` must select the specialization `@m(A,A)`; but `D` is an ambiguous subtype of `A` (an instance of `D` contains two subobjects relative to `A`), so it is not clear which subobject of `d1` and `d2` to chose as actual parameter for `a1` and `a2`.

The check against the occurrence of any of these anomalies is described in the following.

## 4.3 The link-time checks and the calculation of compressed dispatch tables

To achieve a fast dispatch of multimethods, OOLANG uses compressed dispatch tables. Their creation and management is discussed in the following after a short introduction to *non compressed* tables and to their drawbacks.

A non compressed dispatch table associated with a multimethod `@m` is an *n*-dimensional matrix, where *n* is the number of parameters used for the selection of `@m`. Each dimension of the matrix is indexed by the id-s associated to each type (class) in the program. So, when a multimethod is invoked, the *n* id-s associated to the dynamic types of its arguments are obtained and are used to index the dispatch table. The corresponding entry in the table contains a reference

```
class A {…};
class B {…};
class C {…};
class D: public A,
      public B {…};
class E: public C,
      public D {…};

int @m(B,B);
int @m(D,D);
```

inheritance hierarchy and objects layout:

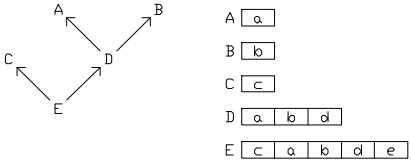

Compressed dispatch structures:

first vector:

| A | B | C | D | E |
|---|---|---|---|---|
| - | P1 | - | P2 | P2 |

second vector:

| A | B | C | D | E |
|---|---|---|---|---|
| - | P1 | - | P2 | P2 |

compressed matrix:

|    | P1     | P2     |
|----|--------|--------|
| P1 | @m(B,B) | @m(B,B) |
| P2 | @m(B,B) | @m(D,D) |

**Figure 13. A sample class hierarchy and multimethods and the related compressed dispatch table**

to the most specific specialization of the multimethod applicable to the *n*-ple of types used for the invocation; this reference is used to invoke the specialization.

This method of dispatching multimethods is very fast because it requires only to access a *n*-dimensional array, but it has the drawback of being very space consuming. In practical cases, fortunately, the tables are sparse and this allows calculating compressed dispatch tables. Compressed tables are obtained by OOLANG using a slight variation of the algorithm in [18].

For each parameter of the multimethod, all the types are analyzed and divided into groups. The reader can refer to [18] for the criteria used in grouping. Every group contains a type that is a supertype of all the other group members; this type is chosen as the representative of the group and is called the *pole* of the group. For example the grouping for the multimethod `@m` in Figure 13 is the following:

- for the first parameter the groups are {B} and {D,E} and the first pole (P1) is B while the second (P2) is D;

- for the second parameter the groups are {B} and {D,E} and the first pole (P1) is B while the second (P2) is D.

Because of loss of transitivity of subtype relation, OOLANG also verifies that no member of the group is an ambiguous subtype of the associated pole.

After the construction of the groups and the election of the poles it is possible to build the tables. To each multimethod `@m` with *n* arguments are associated *n* vectors and an *n*-dimensional matrix (Figure 13).

The *i*-th vector is related to the *i*-th formal parameter. The vectors are indexed with type numbers, i.e. the unique id-s associated to each type in the program (for clarity, however, the type names instead of the type numbers are reported in Figure 13). Thus each vector has size equal to the number of types in the program. The elements in the vectors are the poles. Thus each vector associates each of the types usable for one parameter to the corresponding pole.

The *i*-th dimension of the matrix has size equal to the number of poles associated to the *i*-th parameter. Thus the matrix associates tuples of *n* poles to the most specific applicable methods of `@m`.

At runtime, in an invocation of a multimethod `@m`, for each actual parameter *i*, the *i*-th vector is accessed in order to find the corresponding pole $p_i$. Then the matrix is accessed at co-ordinates $(p_1, …, p_n)$ in order to find the multimethod specialization to call.

This algorithm for table building requires traversing all the types in a program and all the multimethod specializations. Moreover, during the compressed table fill-up, the algorithm verifies that no conflicting specializations are present. So part of the verifications needed at link-time are done for free during the construction of compressed tables.

```
class A {…};
class B: public A {…};
class C: public A {…};
class D: public B, public C {…};

A @m(A,A);
B @m(B,B);
C @m(C,C);
D @m(D,D);
```

**Figure 14. Subtype relation and check of the return type of multimethods**

```
class A {…};
class B: public A {…};

int @m(A,B);
int @m(const A,B);
```

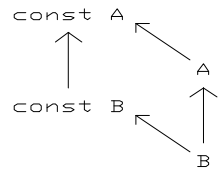

**Figure 15. Example of multimethod with constant parameter and dispatch hierarchy**

The only thing still to be checked is the constraint on the return type: if a specialization is more specific than another the return type of the former must be a subtype of the return type of the latter. During the calculation of compressed tables the specializations are compared to establish which is more specific: the return type check is performed in this phase. So the additional cost introduced by this test is very limited (constant) because the order of the specializations (from the more specific to the less specific) is already obtained for other purposes.

A problem is introduced by the loss of transitivity of the subtype relation. It is possible to notice that it is not sufficient to compare the return type of a specialization with only that of the nearest less specific specialization. Consider the hierarchy in Figure 14: `@m(D,D)` returns `D` that is a subtype of the type returned by `@m(C,C)`. The same way, `@m(C,C)` returns a subtype of the type returned by `@m(A,A)`; so, comparing only couples of neighbour specializations, it seems that the hierarchy is consistent. But `@m(D,D)` can be dynamically invoked when `@m(A,A)` is statically expected and the return type of `@m(D,D)` is not a subtype of the type returned by `@m(A,A)`: it is, in fact, an ambiguous subtype. So, to establish the consistency of a multimethod hierarchy, it is necessary to compare the return type of a specialization with the return type of all the other applicable ones. The algorithm for calculating compressed tables allows this kind of check because it collects all the applicable specializations for every entry of the table and compares them to calculate the most specific. During this phase it is also possible to check if the return type of the most specific one is a non ambiguous subtype of the return types of all the other specializations.

All the reasons above show how the checks needed at link-time to warrant the consistency of multimethod hierarchies are partially done by the algorithm for compressed multimethod dispatch table generation and, for the remaining part, can be integrated into this algorithm without affecting its efficiency. This means that the greatest part of the check can be done for free because is needed by the compressed dispatch table computation algorithm. The complete algorithm for compressed multimethod table generation and verification is presented in [23].

So, at link time, only information about the class hierarchy in the program and the multimethods specializations declared is needed. This information is obtained from the object files.

## 4.4 The interaction of constant arguments with multimethod dispatch

OOLANG supports declaring parameters of functions and multimethods as constant. The presence of constant parameters interferes with the selection of multimethods. Consider the example in Figure 15: the two specializations of the multimethod `@m` are different in that one accepts only constant instances of `A`.

During the selection it is necessary to consider whether an argument is constant or not. So for every type, the selection tables contain two entries, one for the constant version of the type and the other for the non-constant one. This doubles the number of types used for the dispatch; but due to the use of compressed tables, only the size of the vectors doubles while the variation of the matrix size depends on the structure and the number of specializations with constant parameters.

The algorithms presented in [18] are usable even if constant versions of types are considered. In fact, in the example of Figure 15, four types are considered during dispatch: the constant and non-constant versions of `A` and the constant and non-constant versions of `B`. The constant version is considered as a parent of the non-constant version: in fact a multimethod that expects a constant instance can accept a non-constant instance. Instead, if a non-constant instance is expected, it is not possible to accept a constant instance because the body of multimethod will probably need to modify it. At the same time the constant version of *A* is considered as a parent of the constant version of *B*, because the latter class is a subtype of the former (Figure 15).

## 5. Invocation of multimethods: parameter passing and realignment

### 5.1 Parameter passing

Parameters can be passed to an OOLANG function or multimethod in two different modes: by value or by reference. Moreover, a function or multimethod can define a parameter as constant, meaning that the parameter will not be modified by the code of the function.

The passing by reference is internally treated passing only the pointer to the object, as it happens in many other languages.

When a parameter is declared as constant, the OOLANG compiler treats it like a parameter passed by reference for efficiency reasons and according to the fact that a constant object cannot be modified so copying would be useless.

In the passing by value it is necessary to perform a copy of the object on the stack. Because of subsumption it is not possible to know statically the type and the size of the objects passed as parameters. In fact a function accepting a parameter of type *A* can be called passing an instance of *A* or an instance of a class derived from *A* (Figure 16). If only the subobject relative to the expected parameter type were copied, problems could arise when a virtual function or a multimethod is called from inside the static function (as explained in [1]). Moreover it is not possible to know statically the size of the activation record, because the sizes of the function parameters are not known at compile time. For these reasons OOLANG uses a secondary stack to store the copy of the parameters and puts in the primary (normal) stack only the pointers to the objects located on the secondary stack. As calls are nested, in fact, a stack is the proper data structure to keep copies of the parameters. The inefficiency of the double indirection needed to access objects on secondary stack through the pointers that resides on the primary stack is easily optimized by common subexpression elimination [21].

```
class A {...};
class B: public A {...};

int @m(A a) {...}
int @m(B b) {...}

int f(A a) {
    return @m(a);
}

int main() {
    B b;

    return f(b);
}
```

**Figure 16. Interaction between static functions and multimethods in presence of passing by value**

The parameter passing is performed as a cooperative task. The caller pushes the pointers to the objects passed as arguments on the primary stack. The callee, in case of passing by value of non constant arguments, performs the copy of the objects on the secondary stack updating the pointers on the primary stack to point to the copies. This is necessary because, in case of multimethod call, the caller doesn't know which specialization will be called neither if the specialization will actually need to perform a copy of the parameters (this is not necessary for example when the selected specialization of the multimethod requires a constant object).

The copy of parameters on the secondary stack is done through copy constructors that are automatically generated by the OOLANG compiler. The proper copy constructor must be selected at runtime because, as explained, it is not possible to know statically the type of the parameters. In OOLANG the copy constructors are multimethods but traditional virtual functions could be used as well. The same considerations are applicable to the destruction of copies of parameters on exiting the functions.

### 5.2 Realignment of arguments during multimethod invocation

During the analysis of the OOLANG object model (Section 3) was noted that in presence of multiple inheritance not all the subobjects relative to the parents start where the whole object starts (Figure 9).

For this reason, when a multimethod is invoked, it is necessary to realign its arguments, i.e. to modify the pointers passed by the caller to align them to the subobjects statically expected by the multimethod specialization selected.

First realignment vector:

| A | B | C | D | E |
|---|---|---|---|---|
| 0 | 0 | 0 | 0 | Sc |

Second realignment vector:

| A | B | C | D | E |
|---|---|---|---|---|
| 0 | 0 | 0 | 0 | Sc |

Sc = Size of the C subobject

Realignment matrix:

|    | P1   | P2   |
|----|------|------|
| P1 | 0,0  | 0,Sa |
| P2 | Sa,0 | 0,0  |

Sa = Size of the A subobject

**Figure 17. Realignment structures relative to the classes and multimethod in figure 12**

```
class A {…};
class B {…};
class C: public A, public B {…};

B @m(B,B);
C @m(C,C);

int f1(B);

int f2(B b1, B b2) {
    return f1(@m(b1,b2));
}
```

**Figure 18. Realignment of the return type**

The realignment process takes place during multimethod selection using some structures generated by the pre-linker. These structures are similar to those used for the selection (Figure 17). For every parameter $i$ of a multimethod $@m$, besides the vector used for multimethod selection, there is a second vector that contains, for every type, the offset of the subobject relative to the pole associated with the type. So during the selection one vector is used to obtain the id of the pole associated to the type while the other is used to calculate the offset of the pole from the beginning of the object.

Then a second structure is used. It is a matrix with the same number of components and the same size of the dispatch matrix. Every entry is an n-tuple of offsets; these offsets have to be applied too in order to obtain a pointer to the subobject expected by the specialization selected (Figure 17). For example, for a multimethod invocation $@m(b,e)$, where $b$ is of $B$ type and $e$ is of $E$ type, the two pointers $pb$ and $pe$ internally passed by the caller are modified as follows:

```
pb' = pb + 0 + 0
pe' = pe + Sc + Sa
```

Using vectors and tables during multimethod selection it is possible to realign the arguments of the invocation to the types expected by the specialization selected. Moreover these structures allow realizing the realignment in a very fast way.

## 5.3 Interaction between multimethods and static functions

Some problems arise when the object returned by a multimethod is passed to a static function. In fact, often, the object returned by a multimethod is of a subtype of the statically expected type, i.e. the type of the object returned by the specialization that is statically expected to be invoked. But the static function expects the static type of its argument. For example in Figure 18 the invocation $@m(b1,b2)$ is statically expected to return an object of $B$ class and an object of this class is expected as parameter of function $f1()$. But, if $b1$ and $b2$ have dynamic type $C$, the specialization invoked returns an object of the $C$ class; it is necessary to realign this object so that the pointer passed to the $f1()$ function points to the $B$ subobject contained in the $C$ object, a subobject that does not start at the beginning of the whole object.

This example shows how it is necessary to realign the objects dynamically returned by multimethods to the statically expected type. The RTTABLE described in Section 3 is used to obtain the realignment: first, the number of parents is read from the RTTABLE. Then the list of couples parents-offsets is accessed to search the target parent (i.e. the type statically expected) and the offset ($p\_off$) is read. Finally the offset is added to the host object address to compute the address of the subobject relative to the expected parent.

The realignment is necessary wherever the object returned by a dynamically selected function

(multimethod or virtual function as in [22]) is passed to a static function; the object has to be realigned to the statically expected return type.

## 6. Related work

Most of the existing languages that support multimethods are object-based languages. For this reason, they don't face the problems found by OOLANG due to the ambiguity of the subtype relation and to the necessity of object realignment.

Among class-based languages there are CLOS, Polyglot and a variation of Java.

CLOS [16] has been the first language supporting multimethods. It allows the selection of *generic functions* based on the type of multiple parameters. The main feature of OOLANG with respect to CLOS is the ability to perform static type checking.

Polyglot compiles in two phases and checks the return type constraint satisfaction [3] to obtain static typechecking. However it does not support separate compilation and differs mainly from OOLANG under the aspects of the object model and the absence of a pre-linking phase.

Parasitic multimethods [6] are a variation of encapsulated multimethods [8] designed for the Java programming language. They allow separate compilation without link-time checking but loss the symmetry of dispatching because they privilege a receiver argument. Moreover a textual ordering is used to avoid conflict between multimethod specializations.

There are however some object based languages that are interesting for their capabilities related to static typechecking and separate compilation.

Cecil [11] is dynamic, and a static typing can be obtained using particular constructs. In [10] some ideas are described about modular typechecking of multimethods. Cecil does not perform separate compilation, but implements a development environment that keeps the relations among the different parts of the program allowing selective recompilation [9]. Cecil performs heavy optimizations that allow to reduce the costs of multimethods selection thanks to its object oriented optimizer [15].

Dubious [12] allows separate static typing of multimethods. Three type systems are presented allowing different balance between expressiveness and separate compilation. The first type system imposes some limitations on inheritance and multimethod parameters allowing totally modular typechecking; the other type systems relax some constraints but require simple link-time regional checking.

## 7. Conclusions and future work

OOLANG presents the integration of multimethods in an object model similar to the C++ one and allows separate compilation requiring only a link-time check that can be integrated in the algorithm for compressed dispatch table computation, obtaining it efficiently. Moreover, OOLANG addresses the problems of ambiguity and realignment arisen because of the object model adopted.

Now OOLANG is in active development and new features are planned:

- the separation of the subtype relation from the subclass one [5][14]
- the realization of a visual development environment
- the development of an object oriented optimizer